\documentstyle[aps,multicol,epsf]{revtex}
\onecolumn
\begin{document}
\draft
\title{
Nonadiabatic noncyclic geometric phase and ensemble average
 spectrum of conductance
 in disordered mesoscopic rings with  spin-orbit coupling
}
\author{Shi-Liang Zhu and Z. D. Wang$^*$}
\address{ Department of Physics, University of Hong Kong, Pokfulam Road,
Hong Kong, People's Republic of China}

\address{\mbox{}}
\address{\parbox{14cm}{\rm \mbox{}\mbox{}
We generalize  Yang's theory
from the $U(1)$ gauge field to
the non-Abelian $U(1)\times SU(2)_{spin}$ gauge field.
Based on this generalization and
 taking into account the geometric
Pancharatnam phase as well as an effective Aharonov-Bohm (AB) phase
in nonadiabatic noncyclic transport,
we calculate the ensemble average Fourier spectrum
of the conductance
in disordered mesoscopic rings connected to
two leads. Our theory
can explain the experimental results reported
by Morpurgo {\sl et al.}
[Phys. Rev. Lett. {\bf 80}, 1050 (1998)] satisfactorily.
In particular, we clarify that 
the experimentally observed
splitting, as well as some structure on the sides of the
main peak in the 
ensemble average Fourier spectrum, stem from the
nonadiabatic noncyclic Pancharatnam phase and
the effective AB phase, both being dependent on 
spin-orbit coupling.
}}
\address{\mbox{}}
\address{\parbox{14cm}{\rm PACS numbers:  
73.23.-b, 03.65.Bz, 71.70.Ej}}
\maketitle   

As is well known,
the geometric phase~\cite{Aharonov0,Berry,Aharonov,Samuel}
has manifested itself extensively in physics,
particularly in mesoscopic systems where 
quantum interference is extremely
important~\cite{Meir,Loss,Stern,Aronov,Qian}.
Recently, Morpurgo {\sl et al}~\cite{Morpurgo}
reported a novel splitting of the
main peak
(corresponding
to the $hc/e$ Aharonov-Bohm (AB) oscillations)
in the ensemble average Fourier spectrum of the conductance in
open mesoscopic rings~\cite{measured}.
The authors conjectured that the observed splitting is due to the
spin-orbit(SO)-induced Berry's phase~\cite{Loss,Stern,Aronov,Qian}.
It is probably  strong experimental evidence
 showing an important effect of
the SO geometric phase
on  quantum transport. Although it was concluded that,
 in a mesoscopic ring 
possessing the time-reversal symmetry in the absence of AB-flux,
 the SO-dependent
transport can be treated formally in the absence of SO-coupling but with 
an effective magnetic flux~\cite{Meir}, it is unclear yet how to calculate the value of this
flux as well as the ensemble average spectrum of conductance; besides, 
it is not clear either whether the above conclusion is still valid in
the presence of
an arbitrary local magnetic field on the ring(i.e., the aforementioned  
time-reversal symmetry is broken), which appears to be the experimental case in 
~\cite{Morpurgo}.  
Mal'shukov {\sl et al}
 attempted to account for the observed splitting of the main peak,
but were not quite successful~\cite{Mal'shukov}.
Three aspects of the experiment require theoretical
explanation.  First, the
magnitude of the
observed splitting is surprisingly large when compared with an 
estimation based on the adiabatic approximation
in a clean mesoscopic ring. For a clean ring
with radius $r$
subject to a crown-shaped effective magnetic field
${\bf B}_{eff}=(B_0 cos\varphi_r, B_0 sin\varphi_r, B_z)$ in
the cyclindrical coordinates, Stern\cite{Stern} predicted
that the inverse $B_z$ period becomes
$(\Delta B_z)^{-1}=\pi r^2/\phi_0\pm 1/(2B_0)$ at $B_0>>B_z$,
and the splitting would be $\sim 1.2\times 10^{-3} mT^{-1}$
for an estimated experimental parameter $B_0\sim 0.8\ T$,
which is  at least one order of magnitude
less than the observed value.
Secondly, the origin of the side structure on the main peak
needs to be clarified.
 Lastly,  most existing theoretical estimations are crucially based on
the adiabatic or cyclic condition; however, 
neither adiabatic nor cyclic evolution
is well satisfied in the experiment~\cite{Stern,Qian,Wang}.
In view of these facts, we believe that the
{\sl nonadiabatic noncyclic}
geometric phase, essentially similar to the
effective flux addressed formally in Ref.~\cite{Meir},  could play a crucial role in the system,
this  is the key point of the present work,
being essentially different from some existing theoretical
analyses~\cite{Loss,Stern,Aronov,Qian}.
However, it is still highly nontrival to evaluate the
SO-induced geometric phase in the nonadiabatic
noncyclic transport and its effect on
the ensemble average spectrum.

It is worth pointing out that
the relevant geometric phase detected in the experiment
is likely induced by a $U(1)\times SU(2)_{spin}$ field.
In exploring the global geometrical
connotations of  gauge fields
of either the Abelian $U(1)$ type or
the non-Abelian monopole type,
Yang {\sl et al} showed that the nonintegrable
gauge phase factor in the wavefunction gives an
intrinsic and complete description of the relevant field\cite{Yang},
hereafter  refered to as Yang's theory.
In this Letter, we first generalize  Yang's
theory to the non-Abelian $U(1)\times SU(2)_{spin}$ electromagnetic
field. Using a simple one-dimensional(1D) continuum model for
a quasi-1D mesoscopic ring, we then analyze carefully
the nonintegrable phase 
induced by this field
and evaluate its effect
on the splitting
of the main peak in the ensemble average
spectrum of the conductance.
Remarkably, we find that the splitting as well as the side structure
of the main peak
observed by Morpurgo {\sl et al} stem from
the {\sl nonadiabatic noncyclic} geometric phase.

We consider an electron subject to an 
electromagnetic field. The corresponding Hamiltonian
with $U(1)_{em}\times SU(2)_{spin}$ gauge symmetry
is given by\cite{Anandan}
\begin{equation}
\label{Hamiltonian0}
\hat{H}=\frac{1}{2m}({\bf p}+\frac{e}{c}{\bf A}-
\frac{\mu}{c} {\bf a})^2-eA^0+\mu  a^0
+V({\bf r}),
\end{equation}
where $\mu=g\mu_B/2$ with
$g$  the gyromagnetic ratio and $\mu_B=e\hbar/(2mc)$
 the Bohr magneton.
Here $A^\nu=(A^0,{\bf A})$ represents
a $U(1)_{em}$ electromagnetic potential, and
$a^\nu=(a^0,{\bf a})=(-{\bf \stackrel{\rightarrow}{\sigma}}
\cdot{\bf B},{\bf \stackrel{\rightarrow}{\sigma}}\times
{\bf E}/2)$ is an $SU(2)_{spin}$ potential with
${\bf \stackrel{\rightarrow}
{\sigma}}$
denoting  the Pauli matrix.
$V({\bf r})$ is an arbitrary spin-independent
local potential at the point ${\bf r}$.
The Schr\"{o}dinger equation
for the normalized two-component wave function $\Psi (x^\nu)$ reads
\begin{equation}
\label{Schrodinger}
i\hbar\frac{\partial}{\partial t}\Psi (x^\nu)=\hat{H}\Psi (x^\nu).
\end{equation}
By introducing a new wave function
\cite{Anandan},
$\Psi_0(x^\nu)=\hat{U} \Psi(x^\nu)$,
where
\begin{equation}
\label{solution}
\hat{U}=exp(i\frac{e}{\hbar c}\int_\Gamma A_\nu dx^\nu)
\hat{P} exp(i\frac{\mu }{\hbar c}\int_\Gamma a_\nu dx^\nu)
\end{equation}
with $\hat{P}$  the path ordering operator  and $\Gamma$ an integration curve
from a fixed $x_0^\nu$ to $x^\nu$, we find that
Eq.(\ref{Schrodinger}) reduces exactly to 
\begin{equation}
\label{Schrodinger1}
i\hbar\frac{\partial}{\partial t}\Psi_0 (x^\nu)=
\hat{H}_0\Psi_0 (x^\nu)
\end{equation}
with
\begin{equation}
\label{gauge1}
\hat{H}_0=\hat{U}(\hat{H}-i\hbar\frac{\partial}{\partial t})\hat{U}^{-1}
=\frac{(-i\hbar\nabla)^2}{2m}+V({\bf r}).
\end{equation}
Clearly, $\hat{U}$ is a continuous local gauge transformation.
Under this gauge transformation, the Hamiltonian (\ref{Hamiltonian0})
is transformed to a Hamiltonian devoid  of
electromagnetic fields, but with  a phase shift in
the wave function as seen in Eq.(\ref{solution}).
In this sense, the gauge factor in Eq.~(\ref{solution})
is just the
nonintegrable phase in  Yang's theory,
which can
describe completely the $U(1)\times SU(2)_{spin}$
electromagnetic field.
For a mesoscopic ring
where the phase memory is
retained by electrons,
we may conclude that physical properties of the system
in the presence of an electromagnetic field can be
expressed in terms of the same quantity in the absence of
the electromagnetic field, but with 
a nonintegrable phase being taken into account.
An important application is related to the SO coupling:
any spin-independent 
transport quantity can be expressed in terms
of the same quantity in the absence of SO
scattering but with an effective magnetic flux, a fact which
was shown directly by
using the transfer matrix method in a tight-binding form 
for   a mesoscopic ring 
possessing the time-reversal symmetry in the absence of AB-flux~\cite{Meir} .
In fact, with the help of this generalized theory, we are able
to study a disordered mesoscopic
system subject to an electromagnetic field in a simpler way.

We now focus on the phase factor first.
To capture essential physics of geometric
phase in the present quasi-1D system, we employ a simple 1D model.
For a closed path parameterized by arc length $s$,
the total phase factor in Eq.(\ref{solution}) is 
$\gamma_t=\gamma_{AB}+\widetilde{\gamma}$,
where $\gamma_{AB}=2\pi\phi/\phi_0$ is the usual AB phase
with $\phi$ 
the magnetic flux and $\phi_0 =hc/e$,
and $\widetilde{\gamma}$ is the second phase factor
in Eq.(\ref{solution}), which is 
determined by a Schr\"{o}dinger-type equation\cite{Qian,Oh}
\begin{equation}
\label{type}
i\hbar\frac{\partial}{\partial s}|\xi(s)\rangle
=-\mu \stackrel{\rightarrow}{\sigma}\cdot (\frac{1}{v}{\bf B}
-\frac{1}{2c}{\bf \hat{v}}\times {\bf E})|\xi (s)\rangle.
\end{equation}
Here ${\bf \hat{v}}$ is a unit vector 
along the direction of the velocity ${\bf v}=v{\bf \hat{v}}$
and $ds=vdt$.
Equation (\ref{type}) describes the evolution of the spin state
$|\xi\rangle$ governed by the operator
$\hat{U}$. The phase associated with Eq.(\ref{type})
can be further written as $\widetilde{\gamma}  
=\gamma_d+\gamma_{AB}^{eff}+\gamma_p$ \cite{Aharonov} with
the dynamical phase
\begin{equation}
\label{dynamic}
\gamma_d
=\frac{\mu }{\hbar}\int
\langle\xi(s)|\stackrel{\rightarrow}{\sigma}\cdot \frac{1}{v}{\bf B}
|\xi (s)\rangle ds ,
\end{equation}
the effective AB phase
\begin{equation}
\label{effective}
\gamma_{AB}^{eff}=-\frac{\mu }{\hbar}\int \langle\xi(s)|
\frac{1}{2c}\stackrel{\rightarrow}{\sigma}\cdot
({\bf \hat{v}}\times {\bf E})|\xi (s)\rangle ds,
\end{equation}
 and $\gamma_p$ is the Pancharatnam phase, 
to be addressed in detail later.
Here  we emphasize that $\gamma_{AB}^{eff}$
is a kind of geometric phase, though it
seems from Eq.(\ref{effective}) as if
it were a dynamical phase related to
an `effective
magnetic field' $-{\bf v}\times {\bf E}/2c$.
The reason lies in the fact that the two waves
propagating in opposite directions in the ring acquire  phases with
the opposite sign
for $\gamma_{AB}^{eff}$ (simply because it depends on the velocity
direction ${\bf \hat{v}}$),
but the same sign for $\gamma_d$~\cite{Aharonov1}.
The geometrical feature of  $\gamma_{AB}^{eff}$ 
seems to be ignored in some earlier
analyses~\cite{Aronov,Qian}, which appears to be
a minor reason for the existing discrepancy
between  theory and  experiment.
In fact,
$\gamma_{AB}^{eff}$ is just induced by an 
$SU(2)_{spin}$ vector potential ${\bf a}$, and
it is clear from Eq.(\ref{Hamiltonian0}) that
${\bf a}$ plays a  role similar to that of
the $U(1)_{em}$ vector potential ${\bf A}$ in the AB effect.
As a result, it is expected that an effective AB effect can be
induced by this $SU(2)_{spin}$ vector potential~\cite{Wang},
as was also shown by Choi {\sl et al}~\cite{Choi}.

 For a unit vector ${\bf n}=(n_1,n_2,n_3)
=(sin\theta cos\varphi, sin\theta sin\varphi,
cos\theta)$ with  ${\bf n}\in $ a unit sphere $S^2$,
each ${\bf n}$ corresponds to the spin state
$|\xi\rangle=(e^{-i\varphi/2}cos(\theta/2),\ e^{i\varphi/2}
sin(\theta/2))^T$
via the relation
${\bf n}=\langle\xi |\stackrel{\rightarrow}{\sigma}
|\xi \rangle$, where $T$ represents matrix transposition.
The noncyclic Pancharatnam phase accumulated in an evolution of ${\bf n}$
is found to be~\cite{Zhu}
\begin{equation}
\label{spin-phase}
\gamma_p=-\frac{1}{2}\oint_{\partial \Sigma=C}
{\bf n}\cdot d{\bf \Sigma},  
\end{equation}
where $d{\bf \Sigma}$ is an area element on  $S^2$,
$C$ is a specific closed curve on  $S^2$, which
is along the actual path of ${\bf n}(s)$ plus
the shorter geodesic curve from the final point
${\bf n}(s_f)=(sin\theta_f cos\varphi_f, sin\theta_f sin\varphi_f,
cos\theta_f)$ to the initial point
${\bf n}(0)=(sin\theta_i cos\varphi_i, sin\theta_i sin\varphi_i,
cos\theta_i)$.
This Pancharatnam phase can be derived as~\cite{Zhu}
\begin{equation}
\label{phase}
\gamma_p=-\frac{1}{2}\int_{0}^{t_f}\frac{n_1 \dot n_2-n_2 \dot n_1}
{1+n_3}dt
+ arctg\frac{sin(\varphi_f-\varphi_i)}
{ctg\frac{\theta_f}{2}ctg\frac{\theta_i}{2}+cos(\varphi_f-\varphi_i)},
\end{equation}
where $t_f$ is the final time, and ${\bf n}$
is determined by the equation 
\begin{equation}
\label{spin}
\frac{d{\bf n}}{dt}=-\frac{2\mu }{\hbar}({\bf B}-
\frac{1}{2c}{\bf v}\times {\bf E})\times{\bf n} ,
\end{equation}
which represents a spin-$\frac{1}{2}$ particle moving in
an effective magnetic field $({\bf B}-{\bf v}\times {\bf E}/2c)$.
This phase is not equal to the cyclic Aharonov-Anandan (AA)
phase in general~\cite{Aharonov},
but recovers the AA phase
$\gamma_{AA}=-\frac{1}{2}\int_{0}^{\tau}dt(n_1 \dot n_2-n_2 \dot n_1)/
(1+n_3)$ for any cyclic evolution with the
period $\tau$~\cite{Zhu}.
It is remarkable that the nonintegrable
phase in Eq.(\ref{solution}) can be evaluated by simply computing
Eqs.(\ref{dynamic}), (\ref{effective}), and (\ref{phase}); while it is hard
to calculate the value of the effective flux addressed formally in 
Ref.~\cite{Meir}, particularly in the presence of an arbitrary local magnetic
field.

At this stage, as in the experiment~\cite{Morpurgo},
we study a disordered ring
with the Rashba SO-interaction(equivalent to  an internal
electric field ${\bf E}=E{\bf e}_z$),
subject to a local magnetic field
${\bf B}=B_z{\bf e}_z$
and a magnetic flux $\phi=\pi r^2 B_z$. The Hamiltonian,
which is in the form of Eq.(\ref{Hamiltonian0}),
becomes~\cite{Qian,Wang}
\begin{equation}
\label{Hamiltonian-ring}
\hat{H}=\hbar \omega_r [ -i\frac{\partial }{\partial
\varphi_r }+\frac{\phi}{\phi_0}
-\frac{\eta }{2}(\sigma _{x} cos \varphi_r
+\sigma _{y} sin \varphi_r)]^{2}
-\mu  B_z\sigma_z+V(\varphi_r),
\end{equation}
where $\omega_r=\hbar/(2mr^2)$,
$\varphi_r$
is the polar angle, and
the normalized electric field strength $\eta=\mu_B  E r/c\hbar
=2m\kappa r$ with the SO coefficient $\hbar^2\kappa$.

To account for the experimental results naturally,
we investigate the electronic transmission across a disordered
ring connected to external current leads, schematically illustrated
in Fig.3 in Ref.\cite{Buttiker}.
In such a system, the electronic transmission is
significantly affected by the nonintegrable phase. Using
the method originally proposed by
B\"{u}ttiker {\sl et al}\cite{Buttiker} and our generalization
of  Yang's theory,
the transmission coefficient across the ring is found to be
\begin{equation}
\label{transission}
T_g=\frac{\epsilon ^{2}}{b^{4}}
\left| (b-a,1)\widetilde{T}_{+}
[\frac{e^{i\Delta\gamma}}{b^2}
\left(
\begin{array}{cc}
(b^{2}-a^{2}) & a \\ 
-a & 1
\end{array}
\right)
\widetilde{T}_{-}
\left(
\begin{array}{cc}
(b^{2}-a^{2}) & a \\ 
-a & 1
\end{array}
\right)
\widetilde{T}_{+}-\widetilde{1}]^{-1}
\left(
\begin{array}{c}
b-a \\ 
-1
\end{array}
\right) \right| ^{2},
\end{equation}
where $\widetilde{T}_{+}$
and $\widetilde{T}_{-}$
are the transfer matrices of the upper and lower branches of the ring,
$\widetilde{1}$ is the unit matrix,
$a=\pm (\sqrt{1-2\epsilon }-1)/2$,
and $b=\pm (\sqrt{1-2\epsilon }
+1)/2$ with $0\leq \epsilon \leq 1/2$.
$\Delta\gamma=\gamma_{AB}+\gamma_{AB}^{eff}({\bf n}(0))+\gamma_p({\bf n}(0))$
represents the nonadiabatic noncyclic geometrical phase
accumulated in the evolution when electron(with the initial 
spin-state ${\bf n}(0)$) moves one cycle
in the clockwise sense~\cite{dynamic}.
For a beam of electron waves with Fermi wave vector
$k_f$, the rate for electrons to traverse one
round in the ring is $\omega_f=\hbar k_f/(mr)$
for ballistic motion, but is estimated 
approximately to be $\omega_d=l\omega_f/(2\pi r)$
for weak diffusive motion~\cite{Jian-Xin}, where $l$ is the electron mean free path.
This rate can be regarded
as the angular frequency of the otherwise rotating magnetic
field felt by the electron spin~\cite{Jian-Xin},  which is
given by
${\bf B}_{eff}(t)=(B_0^{f,d} cos\omega_{f,d} t,\ B_0^{f,d}
sin\omega_{f,d} t,\ B_z)$ with
$B_0^{f,d}=-\eta\hbar\omega_{f,d}/2\mu$.
Then from the equation $d{\bf n}(t)/dt=-(2\mu/\hbar){\bf B}_{eff}(t)
\times {\bf n}(t)$, ${\bf n}(t)$ is derived exactly as
\begin{eqnarray}
{\bf n}^T(t)
&=&\left ( \begin{array}{lcr}
cos\omega_{f,d} t & -sin\omega_{f,d} t & 0\\
sin\omega_{f,d} t & cos\omega_{f,d} t & 0\\
0 & 0 & 1
\end{array}  \right ) \nonumber \\
\label{crown}
&\times&
\left (
\begin{array}{lcr}
sin^2\chi+cos^2\chi cos\omega_s t 
& cos\chi sin\omega_s t
&\frac {1}{2}sin 2\chi(1-cos\omega_s t)
\\
-cos\chi sin\omega_s t 
& cos\omega_s t
& sin\chi sin\omega_s t 
\\
\frac {1}{2}sin 2\chi (1-cos\omega_s t)
& -sin\chi sin\omega_s t 
& cos^2\chi+sin^2\chi cos\omega_s t 
\end{array}  \right )
{\bf n}^T(0),
\end{eqnarray}
where $\omega_s=\sqrt{\omega_0^2+(\omega_{f,d}+\omega_1)^2}$
and $\chi=arctg[\omega_0/(\omega_{f,d}+\omega_1)]$
with $\omega_0=2\mu  B_0^{f,d}/\hbar$ and $\omega_1=2\mu  B_z/\hbar$.
On the other hand, we can rewrite Eq.(\ref{effective})
clearly as
\begin{equation}
\label{eff-AB}
\gamma_{AB}^{eff}({\bf n}(0))=-\frac{\eta\omega_{f,d}}{2}
\int_0^{2\pi/\omega_{f,d}}
sin\theta cos(\omega_{f,d}t-\varphi)dt
\end{equation}
with $\theta=arctg(\sqrt{n_1^2+n_2^2}/n_3)$ and $\varphi=arctg(n_2/n_1)$.
Substituting Eq.(\ref{crown}) into Eqs.(\ref{phase}) and
(\ref{eff-AB}), the nonadiabatic noncyclic
phases $\gamma_{AB}^{eff}$ and $\gamma_p$  can be 
computed, at least numerically.

For simplicity,  but without loss of generality,
we compute $\widetilde{T}_{+}$ and $\widetilde{T}_{-}$
in a generalized Kronig-Penny ring consisting of $N=N_{+}+N_{-}$
uniformly spaced $\delta -$function barriers with random strengths,
where $N_{+}$ $(N_{-})$ is the number of  barriers
on the upper (lower) branch.
The Hamiltonian for the system 
in the absence of electromagnetic fields reads
$\hat{H}_0=-(\hbar ^{2}/2m)
d^{2}/dx_{\varsigma}^{2}+\sum_{n_{\varsigma}=1}^{N_{\varsigma}}
\lambda _{n_{\varsigma}}\delta
(x_{\varsigma}-n_{\varsigma}a_0)$,
where $\lambda _{n_\varsigma}$ is a  potential strength
parameter, $a_0$ is the lattice spacing, and
$\varsigma=+(-)$ represents the upper (lower) branch. The
spinless electron wave
function in the regions where no potentials are present may be written
as $\psi _{\varsigma}(x_\varsigma)
=A_{n_{\varsigma}}e^{ik_f x_{\varsigma}}+B_{n_{\varsigma}}
e^{-ik_f x_{\varsigma}}$. The coefficients A$_{n_{\varsigma}}$
and B$_{n_{\varsigma}}$ across site $x_{\varsigma}=n_{\varsigma}a_0$
are related through the 
matrix $\widetilde{M}_{n_{\varsigma}}$,
$$
\left(
\begin{array}{c}
A_{\text{(n+1)}_{\varsigma}} \\ 
B_{\text{(n+}1\text{)}_{\varsigma}}
\end{array}
\right) =\widetilde{M}_{n_\varsigma}\left( 
\begin{array}{c}
A_{n_{\varsigma}} \\ 
B_{n_{\varsigma}}
\end{array}
\right)
$$
with
$$
\widetilde{M}_{n_\varsigma}=\left(
\begin{array}{cc}
1-\frac{iV_{n_{\varsigma}}}{2k_f} & -\frac{iV_{n_{\varsigma}}}
{2k_f}e^{-2ik_fn_{\varsigma}a_0} \\
\frac{iV_{n_{\varsigma}}}{2k_f}e^{2ik_fn_{\varsigma}a_0}
& 1+\frac{iV_{n_{\varsigma}}}{2k_f}
\end{array}
\right) ,
$$
where $V_{n_{\varsigma}}=2m\lambda _{n_{\varsigma}}/\hbar ^{2}$
is assumed to be distributed uniformly in an interval
$[-w/2, w/2]$. Then, one can find 
\begin{equation}
\label{T-matrix}
\widetilde{T_{\varsigma}}
=\left(
\begin{array}{cc}
e^{i k_fN_\varsigma a_0} & 0 \\ 
0 & e^{-i k_fN_\varsigma a_0}
\end{array}
\right)
\prod\limits_{n_{\varsigma}=1}^{N_{\varsigma}}
\widetilde{M}_{n_{\varsigma}}.
\end{equation}
Note that $\widetilde{T_{\varsigma}}$ can be further
simplified to a $2\times 2$ matrix~\cite{Ma}.
Substituting the simplified Eq.(\ref{T-matrix}) and $\Delta\gamma$
into Eq.(\ref{transission}), we are able to calculate
the transmission coefficient $T_g$.

For comparison with the experimental observation\cite{Morpurgo},
we plot in Fig.1 the calculated ensemble average Fourier spectrum
of the conductance for unpolarized electrons , which is defined as
\begin{equation}
\label{average}
\langle|G(\nu)|\rangle=\langle|\int_{-B_m}^{B_m}e^{i\nu B_z}
G(B_z) dB_z|\rangle ,
\end{equation}
where $G(B_z)=(e^2/h)\sum_{\pm{\bf n}(0)} {\bar T}_g(B_z)$
with ${\bar T}_g$ the average on the initial spin-orientation\cite{Note},
$\langle\ \rangle$ represents the ensemble average.
Reasonable parameters in the calculation
are determined  as follows.
As in the experiment, $B_m=0.35\ T$,
$v_f\sim 3.0\times 10^5 m/s$, $g\sim14$, 
$r\simeq 1.05 \mu m$
(with $N_{\varsigma}=4200$ and $a_0\sim \frac{\pi}{4}\times 10^{-9} m$),
which leads to
the period in a magnetic field $\simeq 1.2\ mT$;
$w\simeq 0.267k_f$,
which corresponds to the mean free path
$l=96k_f^2a_0/w^2\sim 1.0 \mu m$~\cite{Soukoulis}.
The dimensionless coefficient $\eta\simeq 3.5$,
which corresponds to the experimentally reported
SO coefficient $\hbar^2\kappa
\sim 5.5\times 10^{-10}\ eVcm$\cite{Morpurgo}.
Finally,
it is typical to consider the case $\epsilon=0.25$.
It is worth emphasizing that the essential feature of Fig.1
is  sensitive mainly to the SO coupling parameter $\eta$: no clear
splitting is present in the main peak if $\eta$ is  smaller
than about $1.5$. This
implies that the SO-interaction plays a crucial role
in the splitting.
To clarify the origin of the structure of the main peak,
we plot it under both
adiabatic and nonadiabatic conditions.
The former case is shown in the inset of Fig.1,
where $\gamma_p=\pi
cos \theta_i  (1-cos\chi_a)$ and 
$\gamma_{AB}^{eff}=-\eta\pi cos \theta_i sin\chi_a$
with $\chi_a=arctg(B_0^{d}/B_z)$.
As $(B_z/B_0^{d})$ is no longer small, it is very difficult to have a
clear analytical understanding of
the influences of $\gamma_p$ and $\gamma_{AB}^{eff}$
on the splitting of the peak.
From the inset of Fig.1, we can see that under
the adiabatic approximation a somewhat splitting of the main
Fourier peak is  present  only if we include 
 the effective AB phase;
however, this splitting feature is  obviously not in good agreement with
the experimental result.
After careful analysis, we understand that the effect of
 the adiabatic $\gamma_p$-phase  
is too weak to
play an important role in causing clearly observable splitting.
More remarkably, if ever we take into account both
the Pancharatnam phase and the
effective AB phase 
in the nonadiabatic noncyclic case, as shown by
the solid line in the main
panel of Fig.1, our theoretical result is
in excellent agreement with the
experimental observation, especially for the
splitting and the side structure (two small peaks) of the main peak
(see Fig.5 in Ref.\cite{Morpurgo}).
From Fig.1, we can also see that the Pancharatnam phase plays
a key role in the main splitting, while the two small
side-peaks are closely related to the effective AB phase.
It is therefore clarified for the
first time that the experimentally observed splitting of the
main peak in the
ensemble average Fourier
spectrum stems from the nonadiabatic noncyclic
Pancharatnam phase and the effective AB phase, both being
dependent on the SO coupling.

 Finally, we remark that the radius of the ring  $\sim$ the mean free
path ($\sim 1\mu m$) and the transport is in the weakly diffusive regime
in the experiment. In the present quasi-1D ring, a multi-channel effect, albert weak
and secondary,
may exhibit in the ensemble average spectrum of the conductance (e.g., broadening
and smearing of the peak splitting), which has been ignored in the present work and
may deserve for further detailed study in future. Nevertheless, the effect would
not affect the present conclusion regarding the splitting qualitatively
 because the non-adiabaticity of geometric
phase is unlikely changed significantly in the present weak diffusive ring.

* To whom correspondence should be addressed.
E-mail: zwang@hkucc.hku.hk

{\bf Figure Caption} \newline

Fig.1 The peak of the ensemble average Fourier spectrum
of the conductance
in nonadiabatic noncyclic cases. The
inset shows the corresponding curves under the adiabatic condition.


\begin{references}
\bibitem{measured} As pointed out in Ref.[12], although
the magnetoresistance was
experimentally measured,
the statistical properties of 
conductance and resistance are the same, as
we are interested in the metallic regime.
\bibitem{dynamic} The dynamical phases accumulated in the upper
and lower branches have the same signs, and 
 cancel approximately in a ballistic ring if
the effective magnetic field has cylindrical symmetry.
In the diffuse case,
the phases may cause  conductance fluctuations,
which are not considered here for brevity because our
main concern is the splitting of the main peak.
\end{references}
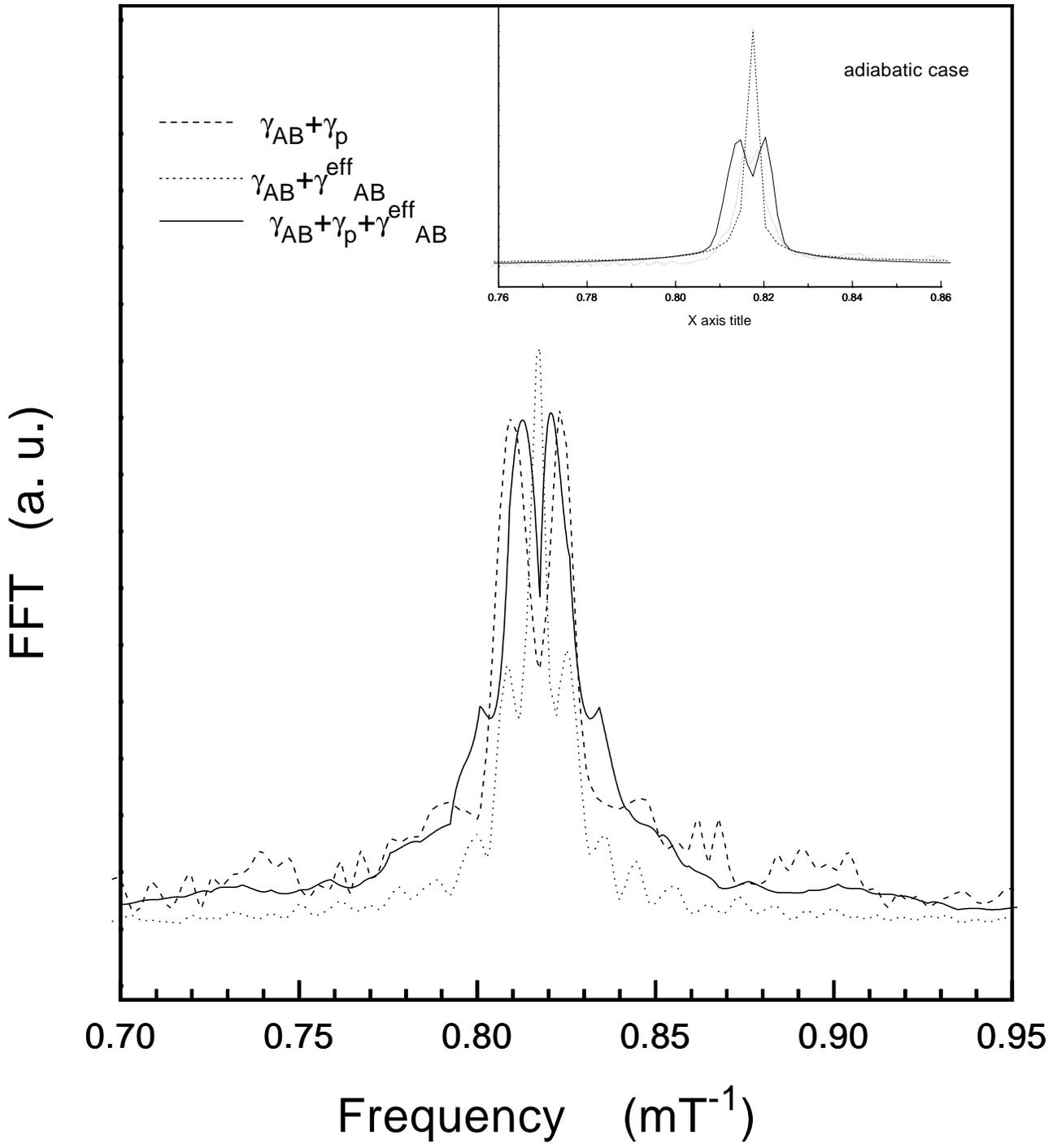
\end{document}